\documentstyle[11pt]{article}
\begin{document}
\begin{titlepage}
\begin{center}
{\Large Theoretical Physics Institute}\\
{\Large University of Minnesota}
\end{center}
\vskip0.5cm
\begin{flushright}
TPI--MINN--93/32--T\\
June 1993
\end{flushright}
\vskip1.cm
\begin{center}
{\huge Valley Singularities and Baryon Number Violation}
\end{center}
\vskip1.cm
\begin{center}
{\large Paolo Provero}
\end{center}
\vskip0.5cm
\begin{center}
{\it Theoretical Physics Institute}\\
{\it University of Minnesota}\\
{\it Minneapolis, MN 55455, USA}\footnote{e--mail: paolo@physics.spa.umn.edu}
\end{center}
\vskip1.cm
\begin{abstract}
We consider the valley--method computation of the inclusive cross section
of baryon number violating processes in the Standard Model. We show
that any physically correct model of the valley action should present
a singularity in the saddle point valley parameters as functions of
the energy of the process. This singularity prevents the saddle point
configuration from collapsing into the perturbative vacuum.
\end{abstract}
\end{titlepage}
Instanton--mediated baryon number violating (BNV) processes in the
Standard Model are suppressed, at zero energy, by the factor
$e^{-2S_I}\sim e^{-400}$, where $S_I$ is the action of the instanton
solution. They were therefore considered as hopelessly unobservable
until, in 1990, Ringwald \cite{ring} and Espinosa \cite{esp} showed
that the cross section grows exponentially when energy is
raised from zero, because of the increase in available phase space
when processes with $O(1/\alpha_W)$ gauge and Higgs bosons in the
final state are considered. This observation raised the hope that
BNV processes could become observable at energies of order of the
sphaleron mass $\sim 10\  {\rm TeV}$.
Despite a lot of theoretical effort (for a review
see Ref. \cite{ma}), a reliable computation of the cross section at the
relevant energies is still lacking.
\par
It is
customary to express the leading term in semiclassical approximation
to the BNV cross section as
\begin{equation}
\sigma_{BNV}(E)\sim\exp \left( 2 S_I F_{hg}(E)\right)
\end{equation}
where $E$ is the energy of the process
and $F_{hg}(E)$ is the so--called holy--grail function.
\par
It is by now well known how to express $F_{hg}(E)$ as a series expansion
in $(E/E_0)^{2/3}$, where $E_0$ is an energy scale of the order of
the sphaleron mass. Choosing
\begin{equation}
E_0=\frac{\sqrt{6}\pi M_W}{\alpha_W}\sim 17\  {\rm TeV}
\end{equation}
the known terms in the expansion of $F_{hg}$ are \cite{scha}
\begin{eqnarray}
F_{hg}(E)&=&-1+\frac{9}{8}\left(\frac{E}{E_0}\right)^{4/3}-
\frac{9}{16}\left(\frac{E}{E_0}\right)^{6/3}\nonumber\\
&&+\frac{3}{32}
\left(4-3\frac{M_H^2}{M_W^2}\right)\left(\frac{E}{E_0}\right)^{8/3}
\log\left(\frac{E}{E_0}\right)+O\left(\frac{E}{E_0}\right)^{8/3}
\label{serie}
\end{eqnarray}
What is needed is a reliable way of extending the computation to
finite values of the ratio $E/E_0$, and therefore predict if BNV processes
can become observable at energies accessible to the next generation of
colliders.
\par
The valley method \cite{kr1,kr2} provides in principle such an analytic
continuation of the expansion (\ref{serie}),
in the (somewhat ill--defined) approximation where initial--state corrections
are neglected.
$\sigma_{BNV}$ is computed,
using the optical theorem, as the imaginary part of the forward elastic
scattering (FES) amplitude of two fermions going into two fermions through
the background field of a deformed instanton--antiinstanton ($I\bar{I}$)
pair. The cross section is expressed, to leading order in semiclassical
approximation, as
\begin{equation}
\sigma_{BNV}(E)\sim {\rm Im} \int dRd\rho_I d\rho_{\bar{I}}
\exp\left(ER-S_v(R,\rho_I,\rho_{\bar{I}})\right)\label{sigma}
\end{equation}
where $S_v(R,\rho_I,\rho_{\bar{I}})$ is the classical action along the
valley trajectory, which connects a far separated $I\bar{I}$ pair
with the perturbative vacuum. The valley trajectory is chosen in
such a way that the functional integration along the directions
orthogonal to it is gaussian, and can therefore be neglected in
the leading order of semiclassical approximation. The trajectory is
parametrized by the distance $R$ between the centers of the (deformed)
$I$ and $\bar{I}$ and their radiuses
$\rho_I$, $\rho_{\bar{I}}$. The integral
is computed in saddle--point approximation.
\par
Unfortunately, the valley action in the Standard Model is not know
beyond an expansion in powers of $\rho/R$ and $\rho$ (from now on
we anticipate that at the saddle point $\rho_I=\rho_{\bar{I}}\equiv\rho$).
The expression of $F_{hg}$ derived from this valley action is the expansion
(\ref{serie}). In general, it is possible to prove \cite{arma}
that the valley method
computation of $F_{hg}(E)$ as an expansion in $E/E_0$ is equivalent
to all orders to the direct computation in the one--instanton sector.
This suggests that $F_{hg}(E)$ can be correctly computed with the valley
method for all energies in which $F_{hg}(E)$ is an analytic function
of $E$.
\par
An approximate model for the electroweak valley action has been proposed
in \cite{kr2}, where the (exactly known) QCD valley \cite{yung}
is supplemented
with a conformal symmetry breaking term
\begin{equation}
S_{csb}=2\pi^2 v^2\rho^2\nonumber
\end{equation}
where $v$ is the VEV of the Higgs field. This means that the $I\bar{I}$
interaction in the Higgs sector is completely neglected, the rationale
being that Higgs boson production in the final state is
subdominant with respect to gauge boson production at $E\ll E_0$.
In this model, the holy--grail function hits the unitarity limit
$F_{hg}=0$ at a finite energy $E_{KR}\sim 45 {\rm TeV}$, and the
cross section loses its exponential suppression. The saddle point
configuration collapses into the perturbative vacuum as $E\to E_{KR}$,
and $F_{hg}(E)$ does not show any singularity \footnote{As already
pointed out in Ref.\cite{grandma}, the singularity found in Ref.
\cite{shve} in the Khoze--Ringwald model is an artifact of a
particular way of solving the saddle point equations.}
between $E=0$ and
$E=E_{KR}$.
\par
In this Letter we will show that the absence of singularities and the collapse
into the perturbative vacuum are artifacts of the approximation defining
the model, and are bound to disappear in a qualitatively realistic
model of the Higgs contribution to the valley action. In fact we will show
that the existence of a singularity preventing the collapse to the
perturbative vacuum is an unavoidable feature of any realistic model
of the valley action, which stems from general features of the valley
trajectory.
\vskip0.5cm
The saddle point evaluation of the integral (\ref{sigma}) gives for the
holy--grail function the expression
\begin{equation}
F_{hg}(E)=ER_*(E)-S_v\left(R_*(E),\rho_*(E)\right)
\end{equation}
where $R_*(E)$, $\rho_*(E)$ are solution of the saddle point equations
\begin{eqnarray}
E&=&\frac{\partial S_v}{\partial R}\label{r}\\
0&=&\frac{\partial S_v}{\partial \rho}\label{rho}
\end{eqnarray}
{}From Eq. (\ref{r}) we can immediately see that the perturbative vacuum
cannot be the saddle point for any nonzero energy.
Indeed, the action must be stationary for all deformations of the fields
around the perturbative vacuum,
and in particular it must be
\begin{equation}
\frac{\partial S_v}{\partial R}\Bigl|_{\rm pert. vacuum}=0
\end{equation}
Note that the perturbative vacuum cannot even be the limit of the saddle
point configurations for $E\to E_{lim}\ne 0$, because it is impossible
to have values of the action arbitrarily close to zero while keeping
$\partial S/\partial R$ finite and nonzero.
\par
It is interesting to see how this seemingly impossible phenomenon occurs
in the Khoze--Ringwald model of Ref. \cite{kr2}. In this model
\begin{equation}
\frac{\partial S_v}{\partial R}=\int d^4x \frac{\delta S_v}{\delta A_\mu^a}
\frac{\partial A^a_\mu}{\partial R} \label{inte}
\end{equation}
and
\begin{equation}
\lim_{E\to E_{KR}}\frac{\delta S_v}{\delta A_\mu^a}=0
\end{equation}
However, the field configuration \cite{yung} is given by
\footnote{We consider here
the concentric $I\bar{I}$ configuration, from which the non--concentric
one can be obtained by a coordinate inversion, see Ref. \cite{yung}.}
\begin{equation}
A_\mu^a(x)=\frac{2}{g} \eta_{a\mu\nu}x_\nu\left\{\frac{1}
{x^2+\rho^2/z}-\frac{1}{x^2+\rho^2 z}\right\}
\end{equation}
where
\begin{equation}
z=1+\frac{\xi^2}{2}+\xi\sqrt{1+\frac{\xi^2}{4}},\ \ \ \ \xi=\frac{R}{\rho}
\end{equation}
so that
\begin{equation}
\frac{\partial A_\mu^a}{\partial R}=\frac{2}{g}\eta_{a\mu\nu}\rho x_\nu
\left\{\frac{1}{\left(x^2 z+\rho^2\right)^2}+
\frac{1}{\left(x^2+\rho^2 z\right)^2}\right\}\frac{dz}{d\xi}
\end{equation}
Being
\begin{eqnarray}
&&\lim_{E\to E_{KR}}\rho_*(E)=0\\
&&\lim_{E\to E_{KR}} z_*(E)=1
\end{eqnarray}
$\partial A_\mu^a/\partial R$ develops a singularity in $x=0$
for $E\to E_{KR}$, which is responsible for the non--vanishing of
the integral (\ref{inte}).
\par
Returning to the general case, we note that the saddle point equations
(\ref{r}) and (\ref{rho}) can be solved to give $R_*(E)$ and
$\rho_*(E)$ whenever
\begin{equation}
W=\det\left(\begin{array}{cc}\partial^2 S_v&\partial^2 S_v\\
\overline{\partial R^2}&\overline{\partial R\partial\rho}\\
\partial^2 S_v&\partial^2 S_v\\
\overline{\partial \rho\partial R}&\overline{\partial\rho^2}
\end{array}\right)\ne 0
\end{equation}
Now for a far--separated $I\bar{I}$ we have $W< 0$, because this is
an unstable (asymptotical)
solution of the equations of motion
\footnote{More precisely, $W\to 0^-$ when $R\to\infty$.}.
This is precisely the reason why a far separated $I\bar{I}$ pair contributes
to the {\em imaginary} part of the FES amplitude
On the other hand,
in the perturbative vacuum, which is a stable solution of the equations
of motion, we have $W>0$. It follows the existence of a boundary in
$(R,\rho)$ space separating the $W>0$ and $W< 0$ regions. On this
boundary $W=0$.
\par
Now consider the line defined in $(R,\rho)$ space by the solutions of
Eq.(\ref{rho}). It joins the far--separated $I\bar{I}$ pair with the
perturbative vacuum, which are both solutions of (\ref{rho}); it follows
that this line will intersect the $W=0$ boundary in a point $(R_i,\rho_i)$.
Here Eqs. (\ref{r}) and (\ref{rho}) cannot be solved, and the saddle point
parameters $R_*(E)$, $\rho_*(E)$ develop a singularity in
\begin{equation}
E_s=\frac{\partial S_v}{\partial R}(R_i,\rho_i)
\end{equation}
\par
The singularity of the field derivative in the Khoze--Ringwald model is
again responsible for the absence of such a singularity, because
our mechanism is based on the fact that the vacuum is a stationary point of
the action. On the other hand, it is not surprising that in the $O(3)$
$\sigma$--model studied in Ref. \cite{grandma}, where the conformal
symmetry breaking term in the action is derived from a true Lagrangian,
a singularity is found in the saddle point parameters:
in this point the saddle
point equations (\ref{r}) and (\ref{rho})
cannot be inverted to give $R_*(E)$ and $\rho_*(E)$
and this {\em implies} that the $W=0$ boundary has been reached (if $W\ne 0$
the saddle point equations can certainly be solved).
\vskip0.5cm
In summary, we have described a general mechanism giving
rise to singularities of the
saddle point valley parameters for any model of the valley action
displaying the correct qualitative behaviour, in the Standard Model as
well as in all the toy models developed to gain insight in the problem
(our argument is independent on the number of parameters needed to
describe the valley trajectory, and is actually even more immediate for
one--parameter models like the two--dimensional Abelian Higgs Model).
This mechanism necessarily prevents the saddle point configuration from
collapsing into the perturbative vacuum. In conclusion, let us remark
the following:
\vskip0.3cm\noindent
(i) Our explanation of the origin of  the singularities
is different from the one proposed in Ref. \cite{doma},
because it is independent from the choice of the weight function defining the
valley trajectory, and relies only on the general features of the valley
trajectory.
\vskip0.3cm\noindent
(ii) Our scenario is too general to give indications about the value of
the holy--grail function at the singularity point: in principle the
singularity can occur after the cross section has lost its exponential
suppression. However, in the $O(3)$ $\sigma$--model of Ref. \cite{grandma},
the singularity occurs, when the parameters of the model are chosen in
a self--consistent way, at an energy where the cross--section is still
exponentially suppressed.
\vskip0.3cm\noindent
(iii) A related problem is the relevance of the FES scattering
amplitude computed in the valley background to baryon number violation
\cite{mshi,ksv,gk1,gk2,gkm}: roughly speaking the problem is that when the $I$
and $\bar{I}$ start to overlap the intermediate states of the FES
amplitude (mapped by unitarity in the final states of the BNV process)
do not have anymore the baryon number predicted by the anomaly law;
the danger is to mistakenly consider baryon number {\em conserving}
contributions to the cross section. Let us notice that in our scenario
the problem becomes less severe, because the complete overlapping of $I$
and $\bar{I}$ is prevented by the singularity.
\vskip0.5cm\noindent
{\bf Acknowledgements}
\vskip0.5cm
The author would like to thank M. Shifman for many useful conversations.
Discussions with R. Guida, K. Konishi, M. Maggiore and N. Magnoli are
also gratefully acknowledged. This work is supported by an INFN grant.

\end{document}